\documentclass{sig-alternate-05-2015}

\usepackage{graphicx,verbatim}

\begin{document}

\setcopyright{acmcopyright}

\doi{10.475/123_4}

\isbn{123-4567-24-567/08/06}

\conferenceinfo{RecSys-DLRS}{September 15, 2016, Boston, MA, USA}

\acmPrice{\$15.00}

\title{Infusing Collaborative Recommenders\\with Distributed Representations}

\numberofauthors{1}

\author{
\alignauthor
Greg Zanotti$^1$, Miller Horvath$^2$, Lucas Nunes Barbosa$^2$\\Venkata Trinadh Kumar Gupta Immedisetty$^1$, Jonathan Gemmell$^1$
    \titlenote{Corresponding Author}\\
    \affaddr{$^1$Center for Web Intelligence}\\
    \affaddr{School of Computing, DePaul University}\\
    \affaddr{Chicago, IL, USA}\\
    \affaddr{$^2$CAPES Foundation}\\ 
    \affaddr{Ministry of Education of Brazil}\\
    \affaddr{Brasilia, DF, BRA}\\
    \email{gzanotti,millerhorvath,lnbarbosa,vimmedis,jgemmell@cdm.depaul.edu}
}

\maketitle
\begin{abstract}

Recommender systems assist users in navigating complex information spaces and focus their attention on the content most relevant to their needs.
Often these systems rely on user activity or descriptions of the content.
Social annotation systems, in which users collaboratively assign tags to items, provide another means to capture information about users and items.
Each of these data sources provides unique benefits, capturing different relationships.

In this paper, we propose leveraging multiple sources of data: ratings data as users report their affinity toward an item, tagging data as users assign annotations to items, and item data collected from an online database.
Taken together, these datasets provide the opportunity to learn rich distributed representations by exploiting recent advances in neural network architectures.
We first produce representations that subjectively capture interesting relationships among the data.
We then empirically evaluate the utility of the representations to predict a user's rating on an item and show that it outperforms more traditional representations.
Finally, we demonstrate that traditional representations can be combined with representations trained through a neural network to achieve even better results.

\end{abstract}



\section{Introduction}
\label{sec:introduction}

Modern websites provide a myriad of conveniences.
Users can browse news articles, watch movies, download books or post their latest vacation pictures.
The sheer volume of available content can be overwhelming and users often suffer from information overload.
Consequently, websites often offer some level of personalization.

By personalizing a user's interaction with the website, the user's attention can be focused on the content most related his or her interests.
Personalization might be achieved by allowing the user to manually select topics of interest.
For example, a visitor to a news aggregation site might select ``soccer'' and ``finance'' from a list of available topics.
Other techniques identify a user's interests by passively observing their activity on the site.

Recommender systems identify relevant content for a user based on interests they have exhibited in the past.
For example, users that have rated popular science fiction movies highly in the past might be recommended the newest sci-fi blockbuster.

Recommendation algorithms often work by computing the similarity between users or between items.
To compute this similarity, a representation is needed.
Users can be represented as a vector over the item space.
Likewise, items are often represented as the users that have consumed them.

Representing users and items is this way has proven to be useful.
The representations are simple to construct and robust when built from large amounts of data.
However, their inherent simplicity means they fail to capture richer semantic relationships.

In this work, we construct alternative user and item representations learned through a neural network.
These resulting representations have several benefits.
First, they are relatively quick to create when compared to other neural network techniques.
Second, they capture meaningful semantic information.
Third, by exploiting this semantic information, recommender systems can better predict a user's interests.

Our experiments on a real-world dataset first offer subjective evidence that distributed representations capture meaningful semantic relationships.
These representations can even be manipulated to produce new representations.
For example, after computing the representation for users, movies, directors, actors and tags in the movie domain, we observe that ``Dr. No'' minus ``Sean Connery'' plus ``Roger Moore'' produces a new representation, most closely related to the representation of ``Octopussy''.
The representations appear to capture the notion that Moore replaced Connery in the Bond franchise.

Our evaluation then turns to the utility of incorporating these representations into a recommender system.
Experiments conducted on the MovieLens 10M Dataset and the Internet Movie Database show that distributed representations provide useful information to the system and improve the root mean square error of predicted ratings.

The rest of this paper is organized as follows.
In Section~\ref{sec:relatedWork} we present related work.
Section~\ref{sec:distributedRepresentations} describes how we learn the distributed representations.
We discuss how we incorporate these representations in a recommender engine in Section~\ref{sec:incorporating}.
Our experimental evaluation is presented in Section~\ref{sec:experimentalEvaluation} and our experimental results in Section~\ref{sec:experimentalResults}.
We conclude the paper with a discussion of our results and direction for future work in Section~\ref{sec:conclusion}.

\section{Related Work}
\label{sec:relatedWork}

Collaborative filtering algorithms work on the assumption that users that have behaved similarly in the past are likely exhibit similar behavior in the future.
User-based collaborative filtering~\cite{konstan1997grouplens,shardanand1995social} works by identifying similar users and recommending items these neighbors have liked. 
Alternatively, item-based collaborative filtering~\cite{deshpande2004item,sarwar2001item} identifies items similar to those the user has liked in the past.
Content-based recommender systems~\cite{balabanovic1997fab, pazzani2007content} model users and items in a content space, such as the terms in a newspaper article or the meta-data of a movie.

In social annotation systems, users described the content of online resources by collaboratively assigning tags.
For example, a user might annotate Dr. No with ``Spy''.
Taken together, the annotations of many users create a rich description of users and items.
In such systems, recommenders can promote items~\cite{Gemmell20121160}, tags~\cite{Gemmell:2010:HTR:1871437.1871543,jaschke2007tag} or even other users~\cite{zhao2010user}.
Tags have also been used in models to predict ratings~\cite{sen2009tagommenders,Zheng20114575} and have been integrated into traditional collaborative filtering~\cite{Tso-Sutter:2008:TRS:1363686.1364171} and content-based recommenders~\cite{deGemmis:2008:ITS:1454008.1454036}.

An alternative way to learn features about items and users is to extract them from usage data. 
Latent features extract via principle component analysis~\cite{jolliffe2002principal} or singular value decomposition~\cite{golub1970singular} have benefited recommender systems by increasing scalability and improving accuracy~\cite{koren2009matrix, paterek2007improving}.

Recent progress in natural language processing and neural networks~\cite{mikolov2013distributed} have shown that it is possible to learn distributed representations of words~\cite{hinton1984distributed} that capture important semantic information.
For example, when trained on the Broadcast News dataset the vector for ``Madrid'' minus ``Spain'' plus  ``France'' is closer to ``Paris'' than to any other learned representation~\cite{mikolov2013efficient,mikolov2013linguistic}.
The notion of capital cities seems to be encoded in the representations.
These techniques have been extended to social networks~\cite{Perozzi:2014:DOL:2623330.2623732}, knowledge graphs~\cite{wang2014knowledge}, sentiment analysis~\cite{zhang2015text} and opinion mining~\cite{irsoy2014opinion}. 

Hybrid recommendation systems~\cite{burke2002hybrid} are a combination of two or more different approaches.
They often improve the accuracy of the recommendations such as when recommending television programs~\cite{BarragansMartinez20104290} or online products~\cite{tran2000hybrid}.

This paper extends upon these previous efforts in recommender systems, distributed representations, and hybrid algorithms. Deep learning and neural networks have been applied to recommender systems before; for example in \cite{dlcf} the authors propose a deep learning-based framework to fuse content and ratings information, while \cite{dlmusic} uses a convolutional neural network to create a content-based music recommender. 
Our model learns distributed representations for users, movies, directors, actors, and tags---contextual information for which representations are useful in their own right (e.g.\ for finding actors that a user may like, or for recommending tags). 
These representations can be used in place of those commonly employed in user-based or item-based recommenders, making it easy to adapt existing recommender systems to our method.
We then leverage the results of the traditional representations with the new representation in a hybrid recommender system.

\section{Distributed Representations}
\label{sec:distributedRepresentations}

\begin{figure*}[t] 
\centering
\includegraphics[width=6.5in]{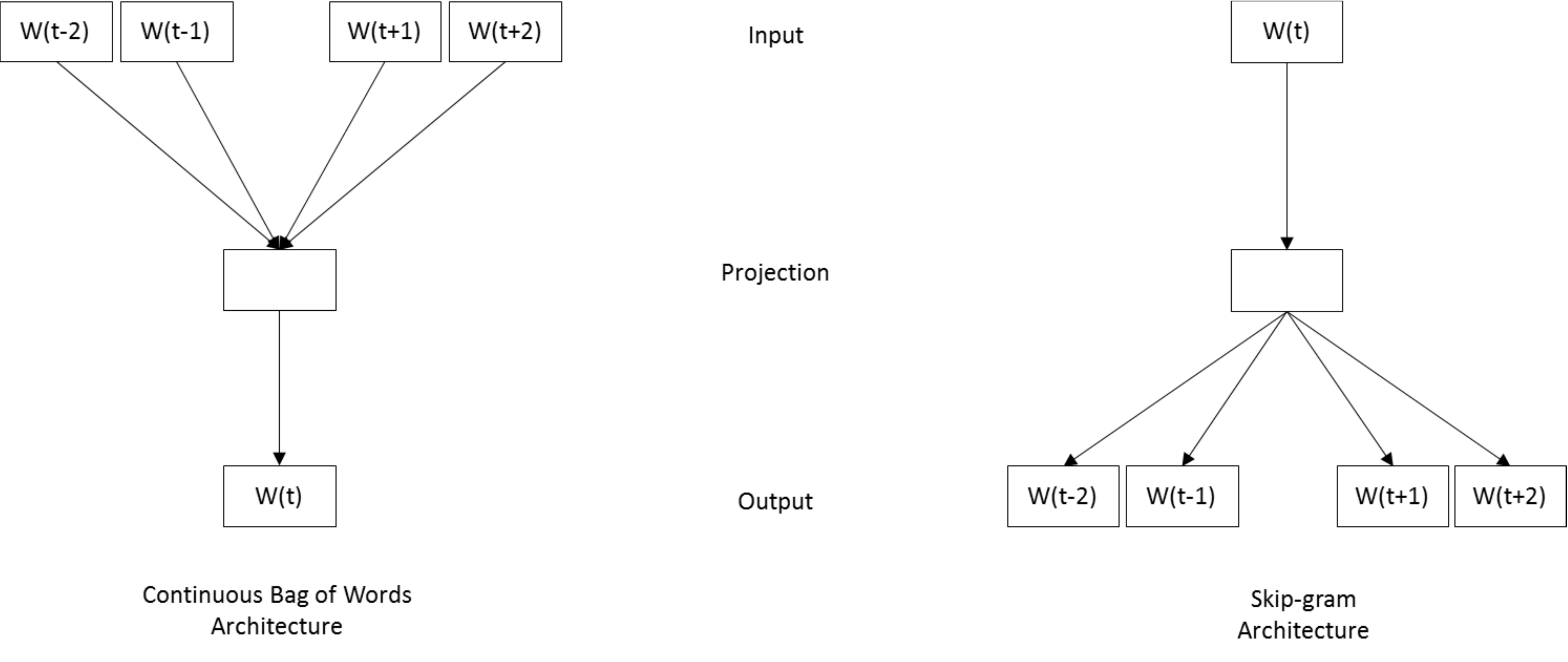}
\caption{Architecture for the Continuous Bag of Word (left) and Skip-gram (right) models.  The first predicts the distributed representation of a word based on its neighbors.  The second predicts the distributed representations of neighbors given a word.}
\label{fig:arch}
\end{figure*}

Distributed representations have a long history of use in machine learning due to their generality and flexibility in representing higher-dimensional data \cite{hinton1984distributed}. The process of creating a distributed representation may add meaningful constructive properties to elements in the transformed vector space as well. Elements from the space can then be used as high-quality lower-dimensional features in other learning problems. 

Recently, developments in continuous space language models have produced effective methods to provide meaningful distributed representations of words and related concepts in the domain of natural language processing. In particular, models utilizing a single hidden layer neural network architecture with a specially designed structure have been shown to be efficient and accurate on a variety of word similarity tasks \cite{mikolov2013efficient, mikolov2013distributed}. These architectures avoid nonlinear hidden layers, thereby significantly reducing training and inference complexity. The distributed representations produced by these models are of relatively low dimension, create meaningful neighborhoods of words, and have useful compositional properties that capture semantic features \cite{mikolov2013linguistic}. Two models in particular are used to produce these representations: the Continuous Bag of Words model and the Skip-gram model.

\subsection{Continuous Bag of Words}

The Continuous Bag of Words model (CBOW) is a neural network composed from an input, projection, and output layer \cite{mikolov2013efficient}. Input words are represented using one-hot encoding, where the dimensionality of the input vector is the number of words in the vocabulary, $V$. The projection layer is of dimensionality $R$, where $R$ is the dimensionality of the target distributed representation. The input to projection layer weight matrix (of dimensionality $V \times R$) is shared across input words, so order does not matter---all words are projected into the same space. The output layer is of dimension $V$.

The training of a CBOW model is optimized to predict the one-hot representation of a \textit{target} word $w_k$ given the surrounding \textit{context} words in a sentence $w_{k-c}$, $...$,  $w_{k-1}$, $w_{k+1}$, $...$, $w_{k+c}$, where $c$ is a fixed window size generally on the order of a small positive integer. Each word in a sentence is sequentially treated as a target word. Multiple sentences comprise the data used to train a CBOW model. The flow of data is described in Figure~\ref{fig:arch}. The one-hot encodings of the context words are averaged and multiplied by the projection matrix to form the hidden layer activations.

The model is trained by optimizing the likelihood function formed by feeding the output layer activations into the softmax function below, where $\textbf{w}_c$ is the collection of context words, and $\sigma (k)$ is the output activation from the $k$th word in the vocabulary:
\begin{equation}
\text{Pr} (w_k | \textbf{w}_c) = \frac{\mathrm{exp}(\sigma (k))}{\sum_{n=1}^V \mathrm{exp}(\sigma (n))}
\end{equation}
The error vector is produced by differencing the one-hot encoding and the vector produced by the softmax function. Training then proceeds by backpropagation, generally using a method like stochastic gradient descent for optimization \cite{backprop, mikolov2013efficient}.

\subsection{Continuous Skip-gram}

The Skip-gram model, also described in Figure~\ref{fig:arch}, is structured in a way similar to the CBOW model, except the inferential task is reversed: given the target word $v_k$, the Skip-gram model is optimized to predict the surrounding context words. The dimensionality of the output layer and input layer are switched, and the hidden to output layer weight matrix shares weights for every output word vector representation. The softmax function acts on the output layer activations for each context word $w_{c,i}$ fed into the network. The softmax functions takes the following form, where the $w_{c,i}$ is the $i$th context word for the target word $w_k$, and $\sigma(c,i)$ is the activation corresponding to $w_{c,i}$:
\begin{equation}
\text{Pr} (w_{c,i} | w_k) = \frac{\mathrm{exp}(\sigma (c,i))}{\sum_{m=1}^V \mathrm{exp}(\sigma (m))}
\end{equation}
The error vector for a set of output activations is computed by summing the difference vectors for each word activation in the output layer. 

\subsection{Efficient Update Methods}
Various schemes have been used to address the significant computational overhead of calculating the updates for each of the output vector representations in the hidden to output weight matrix every time a word is run through a CBOW or Skip-gram model. Two are commonly used in practice: hierarchical softmax and negative sampling \cite{mikolov2013efficient}. Each method has a unique effect on the accuracy of the neural network, and some empirical work has been done as to understanding which one performs best in a given predictive task \cite{mikolov2013distributed}.

Hierarchical softmax uses a binary tree to efficiently represent words in a vocabulary \cite{Morin05hierarchicalprobabilistic, Mnih08ascalable}. Instead of using a matrix of vectors to represent words, each word is instead represented as a path from the root to a unique leaf in the tree, and normalization can be computed in $O (\log V)$ time. Similar words are clustered using simple word-feature algorithms in order to make updates computationally efficient. 

Negative sampling is inspired by noise contrastive estimation \cite{Gutmann:2012:NEU:2503308.2188396}. It deals with the update problem by only updating a sample of the output vectors: the updated word, along with a number of negative samples. A noise distribution, $P_n(w)$ is used to provide the samples for the current word vector representation being updated. An appropriate distribution is chosen empirically; for example, a unigram distribution raised to the power of $\frac{3}{4}$ has been used in practice \cite{mikolov2013distributed}. Updates for backpropagation are derived from the gradients of a loss function based on the noise distribution.

\section{Infusing Recommenders with\\Distributed Representations}
\label{sec:incorporating}

In this section, we propose integrating distributed representations learned from Continuous Bag of Words and Skip-gram models into a collaborative based recommender system.
We begin by presenting the user and item models.
We then present the details of user-based and item-based collaborative filtering.
We propose extensions to these algorithms to incorporate distributed representations.
Finally, we describe how these techniques can be combined using a linear weighted hybrid model.

\subsection{User and Item Models}

There exist many ways to construct user representations in a web application including recency, authority, linkage or vector space models.
In this work we focus on the vector space representation~\cite{salton1975vector} and describe users as:

\begin{equation}
\vec{u} = \langle w(i_1), w(i_2)...w(i_n) \rangle
\end{equation}

\noindent where $w(i_1)$ through $w(i_n)$ represent a user's weights in the vector space.
Such vector space representations offer the opportunity to compute the similarity between users or between items.
In this work, we rely on cosine similarity.

When rated items are used to represent the user, these weights can be taken as the user's ratings for the items.
Similarly, items can be represented as a vector over the user space and the weights are taken as the ratings users have given the item:

\begin{equation}
\vec{i} = \langle w(u_1), w(u_2)...w(u_m) \rangle
\end{equation}

In this work, we compute alternative user and item representations using the Continuous Bag of Words and Skip-gram models.
A user representation takes the same form:

\begin{equation}
\vec{u} = \langle w(f_1), w(f_2)...w(f_{|F|}) \rangle
\end{equation}

\noindent except the weights in the vector space are taken from the discovered representations where $|F|$ is the number of learned features in the representation.
Item representation are drawn in a similar manner.

\subsection{Predicting Ratings}

User-based collaborative filtering (\textbf{UBCF}) is a commonly used recommendation algorithm.
Given a user $u$ and item $i$ a predicted rating $p(u,i)$ is produced by identifying a neighborhood, $N$, of the $k$ most similar users and leveraging their ratings, $r(n,i)$, on the item:

\begin{equation}
p(u,i) = \frac{\sum_n^N sim(u,n) \cdot r(n,i)}{\sum_n^N sim(u,n)}
\end{equation}

The influence of the neighbors in the final prediction is weighted by their similarity to the query user, nearest neighbors receiving more importance.

Item-Based Collaborative Filtering (\textbf{IBCF}) computes the similarity between items rather than between users.
It predicts a user's rating for an item by exploiting ratings that the user gave to similar items:

\begin{equation}
p(u,i) = \frac{\sum_j^J sim(i,j) \cdot r(u,j)}{\sum_j^J sim(i,j)}
\end{equation}

\noindent where $J$ is the set of nearest items to the query item $i$.

We extend user-based and item-based collaborative filtering.
The algorithms work as before except that the user and item representations are substituted with those learned from the Continuous Bag of Words and Skip-gram models.

When computing the similarity between users based on the Continuous Bag of Words representations, we abbreviate the model as \textbf{UBCB}.
When representing users as the representations from the Skip-gram model we abbreviate it as  \textbf{UBSG}.

Likewise, two extensions of item-based collaborative filtering, \textbf{IBCB} and \textbf{IBSG}, are produced by replacing the item representation with those learned in the Continuous Bag of Words and Skip-gram models.

\begin{table*}[htbp]
\centering
\begin{tabular*}{\textwidth}{| @{\extracolsep{\fill}} llll|}
\hline
\textbf{Movies} & \textbf{Directors} & \textbf{Actors} & \textbf{Tags} \\ \hline
Prisoner of Azkaban (0.92)     & Alfonso Cuar\`on (0.71)   & Pam Ferris (0.98)        & best in franchise (0.90)     \\ \hline
Order of the Phoenix (0.82)    & David Yates (0.71)        & Daniel Radcliffe (0.82)  & love this movie so much (0.86)     \\ \hline
Chamber of Secrets (0.79)      & Kazuya Murata (0.67)      & Harry Melling (0.81)     & ignores established character (0.85)     \\ \hline
Half-Blood Prince (0.75)       & Walter Murch (0.66)       & Jason Boyd (0.81)        & potter (0.83)     \\ \hline
Goblet of Fire (0.75)          & Rod Hardy (0.65)          & Richard Griffiths (0.81) & emma thompson (0.82)     \\ \hline

\end{tabular*}
\caption{The top five most similar movies, directors, actors and tags along with their similarity to the distributed representation of the movie ``Harry Potter and the Philosopher's Stone''}

\label{tab:subj_HarryPotter}

\begin{tabular*}{\textwidth}{| @{\extracolsep{\fill}} llll|}
\hline
\textbf{Movies} & \textbf{Directors} & \textbf{Actors} & \textbf{Tags} \\ \hline
Saving Private Ryan (0.70)  & Steven Spielberg (0.79)   & Tom Hanks (0.79)          & best war cinematography (0.70)    \\ \hline
Catch Me If You Can (0.69)  & Lee Unkrich (0.54)        & Russ Meyer (0.59)         & adult diaper commercial (0.70)    \\ \hline
The Terminal (0.67)         & Robert Zemeckis (0.52)    & Rebecca Williams (0.59)   & fun but unrealistic (0.70)        \\ \hline
Forrest Gump (0.60)         & John Lasseter (0.51)      & Stephen Ambrose (0.58)    & tom hanks (0.69)                  \\ \hline
Lincoln (0.57)              & Steve Purcell (0.51)      & Alexander Godunov (0.58)  & fellowship (0.68)                 \\ \hline
\end{tabular*}
\caption{The top five most similar movies, directors, actors and tags along with their similarity to the resulting distributed representation of ``Steven Spielberg'' plus ``Tom Hanks''.}
\label{tab:subj_SpielbergHanks}

\begin{tabular*}{\textwidth}{| @{\extracolsep{\fill}} llll|}
\hline
\textbf{Movies} & \textbf{Directors} & \textbf{Actors} & \textbf{Tags} \\ \hline
Octopussy (0.80)                & John Glen (0.78)              & Roger Moore (0.85)        & desmond llewelyn (0.80)              \\ \hline
For Your Eyes Only (0.78)       & Peter Hunt (0.73)             & Robert Davi (0.77)        & setting circus (0.80)    \\ \hline
A View to a Kill (0.77)         & Michael Damian (0.72)         & Carey Lowell (0.76)       & maud adams (0.80)                    \\ \hline
The Spy Who Loved Me (0.77)     & Jean-Claude Van Damme (0.68)  & Tanya Roberts (0.76)      & kabir bedi (0.80)                    \\ \hline
A Princess for Christmas (0.77) & Harvey Hart (0.66)            & Michael Lonsdale (0.76)   & kristina wayborn (0.80)              \\ \hline
\end{tabular*}
\caption{The top five most similar movies, directors, actors and tags along with their similarity to the resulting distributed representation after computing ``Dr. No'' minus ``Sean Connery'' plus ``Roger Moore''.}

\label{tab:DrNo}
\end{table*}

\section{Experimental Evaluation}
\label{sec:experimentalEvaluation}

In this section, we describe the data used in our experiment.
We then discuss our methodology.
Finally, we present our evaluation metric and baseline models.

\subsection{Data}

The MovieLens 10M dataset~\cite{riedl1998movielens} is made available from GroupLens Research Group.
Released in 2009, it contains 10 million ratings and 100,000 tag applications applied to 10,000 movies by 72,000 users.
After identifying movies in the MovieLens data, we extracted the actors, directors, and other metadata from the Internet Movie Database (IMDb)~\footnote{www.imdb.com}.

In order to generate distributed representations, we first merged these two datasets to create artificial sentences.
The sentences included a user ID, movie ID, set of tags applied by the user to the movie, director, and leading actors.
These sentences were input into the Continuous Bag of Words and Skip-gram models

\subsection{Methodology}

Each user profile was divided equally across five partitions.
Four folds were used as training to build the recommenders. 
The fifth was used to tune the model hyperparameters including the size of the user and item neighborhoods, the length of the distributed representations, the number of epochs to train the representations and the optimal weights for the linear weighted hybrid. The CBOW and Skip-gram models were both trained using Negative Sampling. 

The fifth fold was then discarded, so that the hyperparameters tuned on the fifth fold would not be used to create predictions for it. Four-fold cross-validation was then used with the tuned hyperparameters to create predictions for the remaining folds. The results were averaged across all four folds.

\subsection{Evaluation Metric and Baseline}

The recommenders were evaluated based on their ability to predict a user's affinity for a movie.
Possible metrics include mean absolute error, recall or precision.
In this work, we report results using root mean square error (RMSE).

For each example, in the testing data $T$ consisting of a user $u$, item $i$ and rating $r(u,i)$, we pass the user and item to a recommender which returns a prediction $p(u,i)$.
The RMSE is then computed as:

\begin{equation}
RMSE = \sqrt{\sum_{(u,i) \in T} \frac{(p(u,i) - r(u,i))^2}{n}}
\end{equation}

By squaring the difference between the observation and the prediction, RMSE amplifies and severely punishes large errors.

As baseline models, we use the predictions from the traditional \textbf{UBCF} and \textbf{IBCF} models outlined above. 

\begin{figure*}[t] 
\centering
\includegraphics[width=7in]{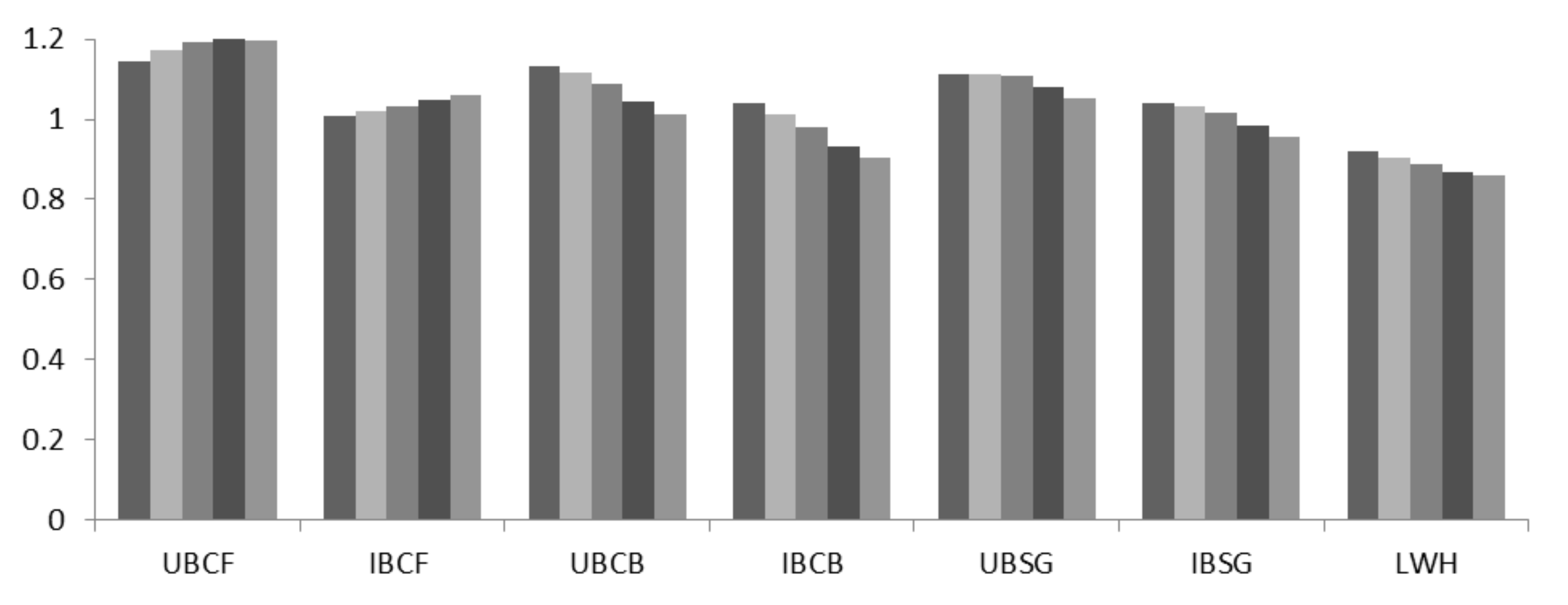}
\caption{Performance of the six recommendation algorithms and the linear weighted hybrid measured as Root Mean Square Error (RMSE) for neighborhoods of size 5, 10, 20, 50 and 100.  The best performing individual recommender was the item-based recommender using the Continuous Bag of Words model (\textbf{IBCB}) with a neighborhood of 100, achieving an RMSE of 0.905.  The linear weighted hybrid was able to reduce the RMSE to 0.858.}
\label{fig:results}
\end{figure*}

\section{Experimental Results}
\label{sec:experimentalResults}

In this section, we present our experimental results.
We begin by presenting subjective evidence that the distributed representation are capturing relevant semantic information.
We then demonstrate empirically that a recommender system infused with the semantic information provided by these distributed representation can produce superior results.

\subsection{Capturing Semantic Information}

By training distributed representations on the MovieLens and IMDb dataset, we hope to capture meaningful relationships between users, movies, directors, actors, and tags.
Consider two similar movies.
They may be similar with regard to the actors and directors of the movie, with regard to the tags users have assigned them, with regard to the users that have watched them, or with regard to all of these dimensions.
If two movies are perceived by users to be similar, then their learned representations should also be similar.

For example in Table~\ref{tab:subj_HarryPotter} we have listed the five most similar movies, directors, actors and tags to ``Harry Potter and the Philosopher's Stone'' using representations learned from the Skip-gram model.
The resulting movies are all from the Harry Potter franchise.
Notice that these movies do not occur together in any training example, since each example contains a single movie.
Alfonso Cuar\`on and David Yates directed most of these movies.
The list of actors include Daniel Radcliffe, the lead actor in the series, as well as supporting actors, Pam Ferris and Harry Melling, whose most popular roles occur in these movies.
The resulting tags include subjective tags such as ``love this movie so much'' which are difficult to judge, but also relevant descriptive tags such as ``potter'' and ``emma thompson''.
Users are able to annotate movies with any tag they wish including actors' names. 

It is also possible to combine distributed representations to produce a new representation.
In this way, representations can be retrieved that share characteristics from both of the originals.
The ability to manipulate representations in this way might enable users to search through information space by selecting multiple movies, directors or actors.

For example, we have added together the representation of the actor Tom Hanks with the director Steven Spielberg.
The result is a new vector of the same dimension.
The most similar movies, actors, directors, and tags are shown in Table~\ref{tab:subj_SpielbergHanks}.
The first three movies were directed by Spielberg and starred Hanks.
As the similarity decreases, we find Forrest Gump starring Hanks and Lincoln directed by Spielberg.
The nearest director to the addition of ``Tom Hanks'' and ``Steven Spielberg'' is not surprisingly Spielberg himself.
Afterward, there is a sharp drop off in the similarity measure from 0.79 to 0.54.
The remaining directors have all worked with Hanks.
Likewise, the most similar actor is Hanks.
Again there is sharp decline to the second most similar actor, 0.79 to 0.59.
The remaining actors worked with either Hanks or Spielberg.
The nearest tags are again subjective (``best war cinematography'') and descriptive (``tom hanks'').

The representations can also be combined in some surprising ways.
Table~\ref{tab:DrNo} presents our final example.
``Sean Connery'' is subtracted from the popular James Bond movie, ``Dr. No''.
Then, ``Roger Moore'', the actor who replaced Sean Connery in the Bond franchise, was added.
The top four resulting movies are Bond films starring Roger Moore.
John Glen and Peter Hunt have both directed Bond films.
Not surprisingly, the representation for Roger Moore is the vector most similar to the new representation.
The remaining actors have all appeared in Bond films.
The resulting tags include the names of actors including Desmond Llewelyn, best known for his role as Q in 17 of the James Bond films.

This subjective evidence suggests that the distributed representations are capturing interesting relationships between the movies, directors, actors, and tags.
It is difficult to quantify this evidence as there is, for example, no definitive similarity between ``Harry Potter and the Philosopher's Stone'' and ``Harry Potter and the Prisoner of Azkaban''.
However, user ratings offer the opportunity to empirically test the quality of the distributed representations.
We now turn our attention to experimental results.

\subsection{Evaluating Predicted Ratings}

Our experimental results are presented in Figure~\ref{fig:results}.
Among the traditional recommender engines we use as our baselines, item-based collaborative filtering (\textbf{IBCF}) performed the best achieving an RMSE of 1.01 when the size of the neighborhood is set to 5.
In both the user-based and item-based recommender systems, smaller neighborhoods produce better results.

When user and item representations are replaced with representations learned through the Continuous Bag of Words model (\textbf{UBCB} and \textbf{IBCB}), the item-based approach again outperforms the user-based approach.
Compared to \textbf{IBCF}, \textbf{IBCB} produces significantly better results with an RMSE of 0.905 when the neighborhood size is set to 100.
In both \textbf{UBCB} and \textbf{IBCB}, it appears a larger neighborhood offers better results.

The Skip-gram representations (\textbf{UBSG} and \textbf{IBSG}) improve upon the traditional approaches, but not as much as the Continuous Bag of Words representations.
Once again, the item-based algorithm outperforms the user-based algorithm achieving an RMSE of 0.956 for a neighborhood of size 100.
Larger neighborhoods appear preferred.

A comparison of these results offers several insights.
First, the distributed representations appear to not only capture semantic information as suggested above, but to represent the interests of users and characteristic of items better than the commonly employed user-based and item-based collaborative filtering techniques.

Second, all item-based approaches outperform their user-based counterpart.
This suggests that, in this domain at least, similarity between items is more informative.
It may be that users watch several different types of movies and consequently create noisy profiles.
In domains where users stick to narrower topics, such as when researchers bookmark journal articles, user-based systems might see an improvement.

Third, traditional approaches prefer small neighborhoods of users while the approaches utilizing the distributed representations prefer larger neighborhoods.
This suggests that traditional user-based and item-based collaborative filtering algorithms struggle to find relevant neighbors and end up introducing noise into the prediction when the neighborhood size is increased.
On the other hand, models incorporating distributed representations appear to benefit from large neighborhoods.
It appears they are able to identify many relevant neighbors.

We now turn out attention to the linear weighted hybrid.
It outperforms all other approaches reducing the RMSE down to as much as little as 0.858.
The hybrid appears able to exploit the strengths of multiple approaches achieving better results than any individual component can achieve alone.

\begin{table}[]
\centering
\begin{tabular}{|l|l|l|l|l|l|l|}
\hline
       & UBCF & IBCF & UBCB & IBCB & UBSG & IBSG \\ \hline
$\alpha$ & 0.032 & 0.058 & 0.144 & 0.419 & 0.122 & 0.225 \\ \hline
\end{tabular}
\caption{The weights assigned to each of the component recommenders indicating their contribution to the linear weighted hybrid with a neighborhood of size 100.}
\label{tab:weights}
\end{table}

The weights associated with the component recommenders are displayed in Table~\ref{tab:weights}.
It is not surprising that the best performing individual recommender, \textbf{IBCB}, receives the most weight, 0.419, in the hybrid.
The second best performing component, \textbf{IBSG}, receives the next highest amount, 0.225.
The two user-based recommenders relying on the learned distributed representation, \textbf{UBCB} and \textbf{UBSG}, receive modest weights.
Little weight is given to either of the two traditional recommenders, \textbf{UBCF} and \textbf{IBCF}.

Several conclusions can be drawn from these observations.
First, since the linear weighted hybrid is trained to optimize the RMSE of the predicted ratings and more weight is given to \textbf{IBCB} than any other component, it seems this model provides the most information.
However, other components garner significant weights as well.
While \textbf{IBCB} is superior to the others, it seems the others still have something to offer.
The contribution of multiple approaches is what allows the hybrid to outperform the individual components.

Second, while it appears the Continuous Bag of Words model is better suited to predict ratings in this domain, the Skip-gram model also outperforms the traditional techniques.
Moreover, since the linear weighted hybrid draws upon the Skip-gram models significantly (0.122 and 0.225), it suggests that the Skip-gram model is learning something that the Continuous Bag of Words model is not.

Finally, since such little weight is given to the models using traditional representations of users and items, it seems that they add little additional information to the hybrid.
In short, the distributed representations contain most of the information of the traditional ones and more.

\section{Conclusions}
\label{sec:conclusion}

In this work, we proposed infusing traditional user-based and item-based collaborative recommender systems with representations learned by applying the Continuous Bag of Words and Skip-gram models to a dataset of users, movies, directors, actors and tags.
We first presented collaborative filtering.
We then presented two approaches for learning distributed representations and described how these representations could be exploited by collaborative recommenders.

Subjective evidence supports the notion that the discovered distributed representation are capturing interesting semantic relationships.
Similar items can be retrieved even though they do not occur together in any training example.
For example, the most similar movie to ``Harry Potter and the Philosopher's Stone'' was ``Harry Potter and the Prisoner of Azkaban''.

A rigorous experiment on a real world dataset demonstrates that the Continuous Bag of Words and Skip-gram representations not only produce better ratings predictions than traditional representations, but appear to subsume them, capturing all the information they do and more.
When including these various techniques in a linear weighted hybrid, even better results can be achieved.

Future work will explore additional datasets.
We are interested in whether these results hold for all domains or if different domains, such as music or news articles, are easier targets for traditional or Skip-gram representations.
Finally, we plan to explore extensions of the Continuous Bag of Words and Skip-gram models themselves to ascertain whether or not the structure of ratings, tagging or content data can be exploited.

\section*{Acknowledgments} 

This work was supported in part by the Science Without Borders Program / CAPES, Coordination for the Improvement of Higher Education Personnel - Brazil.

\bibliographystyle{abbrv}

\end{document}